\documentclass[a4paper,11pt]{article}
\usepackage{jheppub} 
\usepackage{amsmath,amssymb,amsfonts,mathtools,bm}
\usepackage{graphicx}
\usepackage{booktabs}
\usepackage{multirow}
\usepackage{array}
\usepackage{enumitem}
\usepackage{microtype}
\usepackage{xcolor}
\usepackage{float}

\usepackage{placeins}
\usepackage[colorlinks=true,linkcolor=blue,citecolor=blue,urlcolor=blue]{hyperref}
\usepackage[nameinlink,noabbrev]{cleveref}
\graphicspath{{figures/}}
\newcommand{\U}{\mathrm{U}}
\newcommand{\Tr}{\operatorname{Tr}}

\newcommand{\norm}[1]{\left\lVert #1 \right\rVert}
\newcommand{\epsjoint}{\epsilon_{\mathrm{joint}}}
\newcommand{\Dloc}{D_{\mathrm{loc}}}
\newcommand{\Nloc}{N_{\mathrm{loc}}}
\newcommand{\ep}{e_p}
\title{Loop-dependent entangling holonomies in localized topological quartets}
\author[a,b]{Kazuki Ikeda}
\emailAdd{kazuki.ikeda@umb.edu}
\affiliation[a]{Department of Physics, University of Massachusetts Boston}
\affiliation[b]{Center for Nuclear Theory, Department of Physics and Astronomy, Stony Brook University}
\author[c]{Yaron Oz}
\affiliation[c]{School of Physics and Astronomy, Tel-Aviv University,
Tel-Aviv 69978, Israel}
\emailAdd{yaronoz@tauex.tau.ac.il}
\date{}
\begin{document}

\abstract{
A spectrally isolated quartet can admit a local two-qubit description at each point in parameter space and still acquire a loop holonomy outside the local subgroup $\mathrm{U}(2)\otimes \mathrm{U}(2)$. We study this question in three localized topological settings, a BHZ ribbon, a spinful SSH chain, and a BBH corner quartet. On a fixed quartet, changing only the loop can move the holonomy from almost local to entangling. In BHZ, co-rotating and counter-rotating edge-field loops have nearly the same eigenphase data, but only the counter-rotating loop yields an Ising-like entangler. SSH gives a controlled-rotation example in a numerically stable edge quartet. BBH shows the same issue in a higher-order corner quartet. Standard Berry data, including Berry phases, Chern numbers, determinant phases, and eigenphase spectra, do not separate these cases. The main diagnostic is the distance from the loop holonomy to the extracted local subgroup. Canonical two-qubit coordinates are used only after reduction failure has been identified. The quartet is the smallest setting in which this question can be tested explicitly. The same subgroup-reduction problem extends to any isolated multiplet with pointwise product type $D=\prod_{\alpha}d_\alpha$, where the relevant local subgroup is the embedded product group $G_{\mathbf d}=\operatorname{Im}[\prod_{\alpha}\U(d_\alpha)\to\U(D)]$. In the terminology of Ref.~\cite{Ikeda2026}, these examples realize loop-dependent entangling gluing.
}

\maketitle
\flushbottom

\section{Introduction}\label{sec:intro}

An isolated four-state manifold can admit a pointwise two-qubit description and still return from a closed loop by a nonlocal gate. The same quartet may be almost local on one loop and entangling on another. In the BHZ ribbon studied below, co-rotating and counter-rotating edge-field loops act on the same helical quartet, but only the counter-rotating loop leaves the extracted local subgroup by an order-one amount.

Standard reduced Berry data do not resolve this difference. Chern numbers, determinant phases, and Wilson-loop eigenphases record phases or spectra of the loop operator. The question asked here is different. In the extracted local frame, can the microscopic holonomy be written as two independent single-qubit actions? Nearly identical eigenphase data can hide different gate classes, and a large Berry phase can occur on an almost local loop. What matters is whether the holonomy stays inside the extracted local subgroup.

The two-qubit language is the smallest instance of a more general subsystem-reduction problem. If an isolated low-energy multiplet carries pointwise labels of dimensions $(d_1,\ldots,d_r)$, the relevant local structure group is not the full $\U(D)$, with $D=\prod_{\alpha=1}^r d_\alpha$, but the embedded product subgroup

\begin{equation}
G_{\mathbf d}:=\operatorname{Im}\!\left[\prod_{\alpha=1}^r \U(d_\alpha)\longrightarrow \U(D)\right]\subset \U(D),
\qquad (V_1,\ldots,V_r)\mapsto V_1\otimes\cdots\otimes V_r .
\end{equation}

A loop holonomy may preserve these labels point by point and still leave $G_{\mathbf d}$ after parallel transport around a closed path. In that case the loop glues the extracted product structure by an entangling element, or more generally by an element that is nonlocal with respect to the chosen factorization. The real codimension of $G_{\mathbf d}$ inside $\U(D)$ is $D^2-\sum_{\alpha=1}^r d_\alpha^2+(r-1)$. It is positive for every nontrivial factorization. Components transverse to the local algebra are therefore expected unless they are excluded by symmetry, locality, or the choice of controls. The rank-four case $\mathbf d=(2,2)$ is the smallest setting in which this can be examined explicitly and related to standard two-qubit gate classes.

Geometric phases and their non-Abelian generalization are standard \cite{Berry1984,WilczekZee1984}. Holonomies as quantum gates are also well known \cite{ZanardiRasetti1999,Sjoqvist2012}. Wilson-loop and eigenphase diagnostics are widely used in topological band theory and materials calculations \cite{Yu2011,SoluyanovVanderbilt2011,Alexandradinata2014,Gresch2017,BradlynIraola2022}, and non-Abelian Wilson loops are beginning to be measured directly \cite{Sugawa2021,Tyner2024}. The question considered here is narrower. Once a local two-qubit structure has been extracted from a spectrally isolated quartet, does the loop holonomy remain in the corresponding local subgroup, and if not, what gate class does it realize?

The geometric input is the connection on the isolated quartet bundle. Its curvature is the non-Abelian Berry curvature of that bundle, not the Riemannian curvature of the control torus. The control torus may be flat in the coordinates $(\theta_1,\theta_2)$ while the quartet connection still has nonzero curvature. The extra structure needed here is the extracted local subgroup. Once the local frame has been fixed, both the loop holonomy and, infinitesimally, the curvature can be decomposed into components tangent to $\mathfrak g_{\rm loc}$ and components transverse to it. The transverse component is the part that can drive the holonomy outside $\U(2)\otimes\U(2)$.

This question matters in condensed-matter settings where localized boundary multiplets are used as effective qubits or as adiabatic control manifolds. Helical edge states, end modes, and corner multiplets are often described by low-energy labels that factor point by point in parameter space. These labels need not be binary. Spin, edge, valley, layer, orbital, and corner labels can define higher-dimensional product types when the corresponding multiplet remains isolated and the compressed labels remain resolved. The results below show that this pointwise simplicity does not fix the gate class of the loop holonomy. The same localized quartet can support almost local motion, controlled rotations concentrated on one block, or entangling holonomies, depending only on the chosen loop.

The three models play different roles. BHZ shows the contrast most clearly in the present data and ties it to helical-edge physics. SSH isolates the same mechanism in a numerically stable edge quartet. BBH shows that the same question persists in a higher-order corner quartet, although the numerical margins are narrower and the nonlocal response is concentrated on mixed cycles rather than axis cycles.

This work is closely related to Ref.~\cite{Ikeda2026}, which asks when a locally defined subsystem structure can be extended over a twisted family of state spaces. The point needed here is simpler. Once a physically extracted local split has been fixed on a localized quartet, the closed-loop Wilczek--Zee holonomy need not remain local. The fuller discussion of product-state loci, torsor reduction, and Brauer-type questions is given in the Supplementary analyses.

Section~\ref{sec:toymodel} gives a four-state example. Section~\ref{sec:generaldiagnostics} states the rank-$D$ subgroup-reduction criterion. Section~\ref{sec:diagnostics} introduces the extracted local frame and the diagnostic $\Dloc$. Sections~\ref{sec:bhz}, \ref{sec:ssh}, and \ref{sec:bbh} present the three microscopic models. Section~\ref{sec:Berry} explains why standard Berry data do not resolve the distinction, and Section~\ref{sec:discussion} closes with the common gate geometry and the experimental outlook.

\section{Illustrative example}\label{sec:toymodel}

A four-state model makes the issue transparent. Consider a Hilbert space
$\mathcal H \simeq \mathbb C^2_A \otimes \mathbb C^2_B$
with fixed product basis
$\{ |00\rangle,\ |01\rangle,\ |10\rangle,\ |11\rangle \}$.
At each point in parameter space the system admits a local two-qubit description, in the sense that states can be read as products of two effective binary labels. This is the pointwise factorization used later in the microscopic models. The question is whether that local tensor-product structure is preserved after adiabatic evolution around a closed loop.

To illustrate the possibilities, consider three Hermitian generators acting on $\mathcal H$:
\begin{eqnarray}
K_{\mathrm{loc}} &=& \phi\, I \otimes Z,\nonumber\\
K_{\mathrm{ctrl}} &=& \phi\, P_0 \otimes Z, \qquad P_0 = \frac{I+Z}{2},\nonumber\\
K_{\mathrm{ent}} &=& \phi\, Z \otimes Z,
\end{eqnarray}
where $Z$ is the Pauli matrix and $\phi$ is a real parameter.
The corresponding unitaries are $U = e^{-iK}$.
These three cases represent qualitatively distinct types of transport. (i)
$K_{\mathrm{loc}}$ generates a local operation acting only on the second qubit,
$U_{\mathrm{loc}} = I \otimes e^{-i\phi Z}$,
which lies in the local subgroup $U(2)\otimes U(2)$. (ii) $K_{\mathrm{ctrl}}$ generates a controlled rotation, in which one qubit acts as a
control for the other. This operator is not of the form $A\otimes B$, even though it
acts nontrivially only on a single block, i.e. the action on the second qubit is conditioned on the state of the first qubit. (iii) $K_{\mathrm{ent}}$ generates an Ising-type entangling gate, which couples the two
qubits and cannot be decomposed into independent single-qubit actions.
Thus, even within a fixed two-qubit Hilbert space, there exist transformations that
preserve the pointwise tensor-product structure but are globally nonlocal.

Local and entangling generators can share the same eigenvalue structure. For example,
$I \otimes Z$ and $Z \otimes Z$ have the same eigenvalue multiset $\{\pm \phi,\pm \phi\}$.
Quantities that depend only on eigenvalues, such as eigenphase spectra or determinant phases, cannot distinguish a local transformation from an entangling one. The microscopic examples below show the same phenomenon for loop holonomies with nearly identical eigenphase data.

The same point can be stated geometrically. Suppose parameter space is covered by overlapping patches $U$ and $U'$, each equipped with a local product basis. On the overlap, the two descriptions are related by a transition operator $X$:

\begin{equation}
|\psi\rangle_{U'} = X^{-1} |\psi\rangle_U \ .
\end{equation}

If $X \in U(2)\otimes U(2)$, the product structure is preserved globally. If $X$ lies outside this subgroup, a state that appears as a product in one patch may appear entangled in another. This is the mechanism referred to here as entangling gluing. In the microscopic setting, the role of $X$ is played by the loop holonomy $U(C)$.

The same gluing statement has a direct higher-dimensional form.  For a bipartite local model $\mathcal H_{\rm loc}\simeq\mathbb C^{d_A}\otimes\mathbb C^{d_B}$, local generators have the form $X_A\otimes I+I\otimes X_B$ up to an overall phase.  By contrast,
\begin{equation}
K_{\rm block}=\sum_a P_a\otimes h_a,
\qquad
K_{\rm int}=Q_A\otimes Q_B,
\end{equation}
with at least two different $h_a$ or with both $Q_A$ and $Q_B$ non-scalar, generically lies outside $\mathfrak u(d_A)\oplus\mathfrak u(d_B)$ even though the two factors are well defined at each point.  For $r$ factors, the corresponding local Lie algebra is
\begin{equation}
\mathfrak g_{\mathbf d}=\left\{i\alpha I+
\sum_{\beta=1}^r I_{d_1}\otimes\cdots\otimes X_\beta\otimes\cdots\otimes I_{d_r}: X_\beta\in\mathfrak{su}(d_\beta)\right\}.
\end{equation}

This obstruction is not specific to two qubits. The two-qubit quartet is only the smallest case that can be displayed without the extra bookkeeping required for qudits or multipartite factors. Cartan coordinates, concurrence, and two-qubit entangling power enter later only as convenient labels for this minimal case, not as part of the general definition.

The example above shows three points. A pointwise local factorization does not guarantee local return after a closed loop. Different loops on the same four-state manifold can realize different gate classes. Eigenphase data alone do not determine locality. The BHZ ribbon, SSH chain, and BBH model give microscopic instances of these possibilities, with co-rotating loops that stay almost local, single-edge loops that act as controlled rotations, and counter-rotating or mixed loops that become entangling. The next section introduces the diagnostics used to distinguish these cases.

\section{Diagnostics of local and entangling holonomy}\label{sec:diagnostics}

\subsection{General subsystem formulation and the quartet benchmark}\label{sec:generaldiagnostics}

Let $\mathbf d=(d_1,\ldots,d_r)$ be a proposed local factorization type and let $D=\prod_{\alpha=1}^r d_\alpha$.  The general construction starts from an isolated rank-$D$ spectral projector
\begin{equation}
P_D(\lambda)=\sum_{a=1}^{D}|u_a(\lambda)\rangle\langle u_a(\lambda)|,
\end{equation}
together with a fixed set of compressed observables whose spectra resolve the intended product labels.  When these compressed observables remain sufficiently compatible and split the isolated multiplet into the prescribed blocks, they define a pointwise frame, after choosing the ordered product basis of $\bigotimes_{\alpha=1}^r\mathbb C^{d_\alpha}$,
\begin{equation}
F(\lambda):\mathbb C^D\longrightarrow P_D(\lambda)\mathcal H_{\rm micro}.
\end{equation}
The closed-loop holonomy $U(\mathcal C)\in\U(D)$ then has a local interpretation only if it lies in the embedded product subgroup
\begin{equation}
G_{\mathbf d}:=\operatorname{Im}\!\left[\prod_{\alpha=1}^r\U(d_\alpha)\longrightarrow\U(D)\right]\subset\U(D),
\qquad (V_1,\ldots,V_r)\mapsto V_1\otimes\cdots\otimes V_r,
\end{equation}
up to the discrete relabelings of identical factors when those relabelings are physically allowed.  The dimension-independent subgroup-distance diagnostic is
\begin{equation}
D_{\mathbf d}(\mathcal C)=\min_{V_\alpha\in\U(d_\alpha)}\left\|U(\mathcal C)-V_1\otimes\cdots\otimes V_r\right\|_F .
\end{equation}
It vanishes exactly when the loop transport preserves the extracted product decomposition.  Since the local subgroup is a lower-dimensional target inside $\U(D)$, transverse connection components are expected unless excluded by symmetry or by the choice of controls.  Infinitesimally, the same statement is obtained by projecting the Berry curvature onto the complement of $\mathfrak g_{\mathbf d}$, so a nonzero transverse component is the local source of higher-dimensional subsystem-nonlocal holonomy.

Below we evaluate the construction for $\mathbf d=(2,2)$. This is the smallest nontrivial choice. A single factor has no entangling subgroup to leave, while the rank-four case already supports controlled rotations and Ising-type entanglers. At this dimension the subgroup test can be supplemented by Cartan coordinates, two-qubit entangling power, and concurrence witnesses. For higher-dimensional multiplets these secondary diagnostics would be replaced by $D_{\mathbf d}$, operator-Schmidt data across the chosen partitions, generalized entanglement witnesses, and process tomography on $\bigotimes_\alpha\mathbb C^{d_\alpha}$. The reduction question itself is unchanged.

\subsection{Entangling holonomy}
Here we do not assume a unique tensor-product structure a priori. Instead, we ask whether a physically motivated factorization, extracted from compressed observables whose quartet spectra split into two doublets, can be continued around a loop without leaving the associated local subgroup. The underlying observables need not be binary in the full Hilbert space; binary labels emerge only after compression to the quartet. The observable pair $(O_A,O_B)$ is fixed once per model from locality and symmetry considerations and is not tuned loop by loop. Specifically, we build the rank-four projector
\begin{equation}
P_4(\lambda)=\sum_{a=1}^4 |u_a(\lambda)\rangle\langle u_a(\lambda)|
\end{equation}
from the four low-energy states nearest zero energy, compress the two observables into that subspace, and use them to extract a pointwise local frame. The extracted local frame is used only when the quartet remains well isolated and the compressed observables are sufficiently compatible, as verified by the quartet gap $\Delta_4$ and the normalized commutator
\begin{equation}
\epsjoint(\lambda)=\frac{\norm{[P_4 O_A P_4,\,P_4 O_B P_4]}_F}{\mathrm{range}(P_4 O_A P_4)\,\mathrm{range}(P_4 O_B P_4)}.
\end{equation}
For a closed loop $\mathcal C$, the extracted Wilczek--Zee holonomy $U(\mathcal C)$ is then compared with the local subgroup through
\begin{equation}
\Dloc(\mathcal C):=\min_{A,B\in\U(2)} \norm{U(\mathcal C)-A\otimes B}_F.
\end{equation}

This equation is the $\mathbf d=(2,2)$ specialization of $D_{\mathbf d}$. All statements about reduction failure rely on this subgroup-membership layer. The later gate-class labels use the special structure of two qubits only after that test has been applied.

The question is whether the microscopic holonomy can be written as two independent single-qubit actions. This is a subgroup-membership problem rather than a spectral one. The Frobenius distance to the local subgroup $U(2)\otimes U(2)$ is a basis-covariant diagnostic of this property. It vanishes exactly when the holonomy lies in that subgroup. Thus $D_{\mathrm{loc}}(C)=0$ means that the holonomy is local, a small value means that it is almost local, and a finite value means that it has left the local subgroup. We use $D_{\mathrm{loc}}$ as the main diagnostic of local-reduction failure. It is not an entanglement monotone and not a complete classifier of gate class.

Once reduction fails, one can ask which nonlocal gate class has appeared. For that purpose we use canonical Cartan coordinates, exact two-qubit entangling power $\ep$, Schmidt spectra, and fitted effective generators. These quantities describe how the nonlocal holonomy sits inside the two-qubit gate geometry after $\Dloc$ has shown that the loop has left the local subgroup.

This order also explains the presentation of the models. BHZ is discussed first because the loop contrast is sharpest there in the present data. SSH then isolates the same mechanism in the numerically most stable setting. BBH extends the same question to a higher-order corner quartet. In all three cases the observable pair is fixed model by model and is not tuned loop by loop.

\subsection{Curvature, homotopy and subgroup content.}
Let $A=F^\dagger dF$ denote the anti-Hermitian connection in the extracted quartet frame.  The associated curvature
\begin{equation}
F_A=dA+A\wedge A
\end{equation}
is the usual non-Abelian Berry curvature.  It is defined on the control torus, but it is not a curvature of the torus as a Riemannian manifold.  The present tori are flat parameter spaces; the nontrivial object is the gauge curvature of the quartet bundle over them.  For a small contractible boundary $\partial\Sigma$, the standard curvature expansion gives, up to path-ordering corrections at higher order,
\begin{equation}
U(\partial\Sigma)=I-\int_\Sigma F_A+O({\rm area}^{3/2}).
\end{equation}
Thus a null-homotopic loop can acquire a nontrivial holonomy whenever the quartet connection is not flat.  The original diagnostic used here begins after this standard step: in the extracted local frame we project the curvature onto the complement of the local Lie algebra,
\begin{equation}
F_A^{\perp}:=(1-\Pi_{\rm loc})F_A,
\end{equation}
and regard $F_A^{\perp}$ as the infinitesimal source of nonlocal closed-loop holonomy.  A contractible loop is therefore not expected to be automatically local; it is local only when the integrated nonlocal component is absent or cancels over the disk.

The same distinction clarifies the role of the fundamental group.  The group $\pi_1(T^2)\simeq\mathbb Z^2$ labels homotopy classes of closed curves, and for a flat connection the holonomy would reduce to a representation of this group.  The quartet connections studied here are not flat in this sense, so the holonomy is not determined by the homotopy class alone (Figs.~\ref{fig:contractibleloops} and \ref{fig:contractibleradius}).  Noncontractible cycles can differ because they sample different components and orderings of the connection, while contractible cycles can still be nonlocal through $F_A^{\perp}$.  Orientation reversal is a separate statement: for any fixed loop $C$, one has $U(C^{-1})=U(C)^\dagger$, so $\Dloc$ and $\ep$ are unchanged even though the generator changes sign.

\section{BHZ ribbon: one quartet, distinct reduction outcomes}\label{sec:bhz}

\subsection{Co-rotating and counter-rotating loops on one quartet}

The BHZ ribbon gives the sharpest loop contrast in the present data. On the same helical quartet, co-rotating edge fields remain almost local, while counter-rotation gives an entangling holonomy with nearly the same eigenphase data. This makes the comparison with standard Berry diagnostics especially direct.

We study a ribbon slice at fixed $k_x=0$ with open $y$ direction, independently rotating top and bottom edge Zeeman fields, width $L_y=10$, and parameters $(M,B,A,\lambda_R,h_e)=(1,1,1,0.2,0.6)$ \cite{BHZ2006,HasanKane2010,QiZhang2011}. The control torus is spanned by the edge-field angles $(\theta_T,\theta_B)$.

We write $\tau_\mu$ for the orbital Pauli matrices and $\sigma_\mu$ for spin, with $\tau_0=\sigma_0=I_2$. With $c_y(k_x)$ the four component row spinor, the Hamiltonian is
\begin{equation}
\label{eq:Ham_BHZ}
H_{\rm BHZ}(k_x,\theta_T,\theta_B)
=
\sum_{y=1}^{L_y} c_y^\dagger h_0(k_x)c_y
+
\sum_{y=1}^{L_y-1}\bigl(c_y^\dagger T_y c_{y+1}+\text{h.c.}\bigr)
+
 c_1^\dagger V_T(\theta_T)c_1
+
 c_{L_y}^\dagger V_B(\theta_B)c_{L_y},
\end{equation}
with
\begin{align}
h_0(k_x)
&=
\bigl(M-4B+2B\cos k_x\bigr)\tau_z\otimes\sigma_0
+
A\sin k_x\,\tau_x\otimes\sigma_z
+
\lambda_R\sin k_x\,\tau_x\otimes\sigma_y,\\
T_y
&=
B\,\tau_z\otimes\sigma_0
-
\frac{iA}{2}\tau_y\otimes\sigma_0
+
\frac{i\lambda_R}{2}\tau_x\otimes\sigma_x,\\
V_T(\theta_T)
&=
h_e\,\tau_0\otimes(\cos\theta_T\,\sigma_x+\sin\theta_T\,\sigma_y),\\
V_B(\theta_B)
&=
h_e\,\tau_0\otimes(\cos\theta_B\,\sigma_x+\sin\theta_B\,\sigma_y).
\end{align}
At $k_x=0$ the $\sin k_x$ terms in $h_0$ vanish, but the $y$ hopping still contains the Rashba term proportional to $\lambda_R$. The annulus calculation in the next subsection keeps the same Hamiltonian and varies $k_x$.

The compressed observables are ribbon side and a spin-derived label,
\begin{equation}
O_A^{\rm BHZ}=y,
\qquad
O_B^{\rm BHZ}=s_z.
\end{equation}
Inside the quartet, the first effective qubit is the edge label and the second is the binary label extracted from compressed $s_z$. Although $s_z$ is not conserved once $\lambda_R\neq0$, its compression into the isolated quartet remains well split over the benchmark window and provides a stable helical pseudospin label. At $k_x=0$ this second factor coincides with the Kramers pair; away from $k_x=0$ we refer to it as the extracted helical pseudospin. We denote the corresponding Pauli operators by $Z_{\rm edge}$ and $Z_h$, respectively.

\begin{figure}[H]
  \centering
  \includegraphics[width=\textwidth]{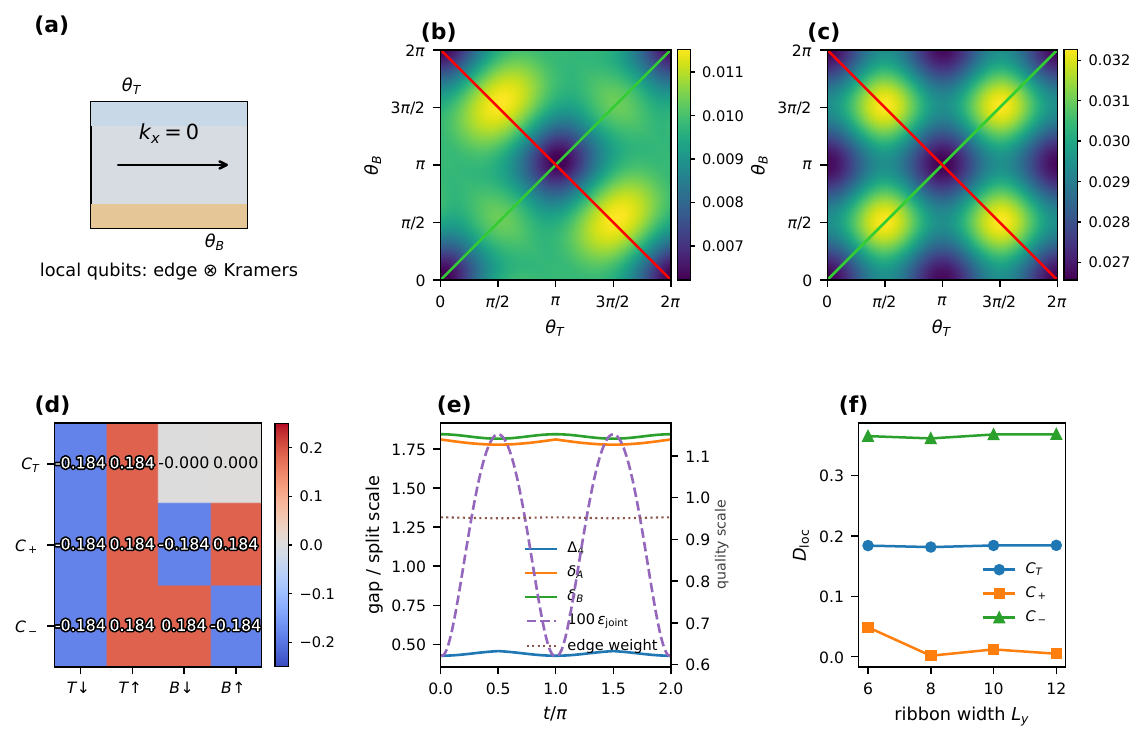}
  \caption{BHZ ribbon benchmark. (a) Ribbon slice with independent top and bottom edge angles $(\theta_T,\theta_B)$. (b) Pointwise $\epsjoint$ map on the edge-angle torus. (c) Link-level distance to the local Lie algebra. (d) Phase pattern of the counter-rotating holonomy in the extracted basis $[T\downarrow,T\uparrow,B\downarrow,B\uparrow]$; at $k_x=0$ these arrows coincide with the extracted helical pseudospin basis, and the panel label $K$ denotes that second qubit. (e) Along-loop quality on $C_-$. (f) Width dependence: the counter-rotating response remains finite within the benchmark family while the co-rotating control stays near local.}
  \label{fig:bhz}
\end{figure}

The quartet remains well isolated over the full edge-angle torus: $\Delta_4^{\min}=0.43$, $\max\epsjoint=0.01$, and the weight in the outer two rows stays above $0.95$. The fixed-slice results are summarized in \Cref{fig:bhz}. For the symmetry-defined loops
\begin{equation}
C_T:(\theta_T,\theta_B)=(t,0),\quad
C_B:(0,t),\quad
C_+:(t,t),\quad
C_-:(t,-t),
\qquad t\in[0,2\pi],
\end{equation}
the response is
\begin{equation}
\Dloc(C_T)=\Dloc(C_B)=0.18,
\qquad
\Dloc(C_+)=0.01,
\qquad
\Dloc(C_-)=0.37.
\end{equation}

The subscripts in $C_+$ and $C_-$ label relative edge winding rather than opposite orientations of one loop.  In $\pi_1(T^2)$, $C_+\sim(1,1)$ and $C_-\sim(1,-1)$, whereas the orientation reverse of $C_+$ is $(-1,-1)$.  Thus $C_-$ is not $C_+^{-1}$; it is a different primitive cycle that reverses only the bottom-edge field relative to the top-edge field.  For any fixed loop $C$, by contrast, reversing its traversal gives $U(C^{-1})=U(C)^\dagger$ and leaves $D_{\rm loc}$ and $e_p$ unchanged.

On this torus, the same helical quartet supports co-rotation that stays almost local, single-edge cycles with intermediate $\Dloc$, and a counter-rotating entangler. As summarized in \Cref{tab:cartan}, the co-rotating and counter-rotating loops also share the same sorted microscopic eigenphase quadruplet to the reported precision, so the contrast is not explained by eigenphase size alone. At the fitted-generator level this equality is exact because $I\otimes Z_h$ and $Z_{\rm edge}\otimes Z_h$ share the same eigenvalue multiset $\{\pm\phi,\pm\phi\}$. The microscopic loops inherit this equality up to small higher-order corrections.

The mechanism is visible in the $4\times4$ low-energy structure. Because the helical handedness reverses between the top and bottom edges, the same in-plane field rotation enters the two edge blocks with opposite helical sign. Co-rotation therefore reduces to a common action on both edge sectors and collapses to an almost local $I\otimes Z_h$ response, whereas counter-rotation converts that sign reversal into a relative phase between the two edge blocks and leaves a residual $Z_{\rm edge}\otimes Z_h$ term. Writing $P_T=(I+Z_{\rm edge})/2$, the fitted generators follow exactly this pattern:
\begin{equation}
K_T\approx \phi\,P_T\otimes Z_h,
\qquad
K_+\approx \phi\,I\otimes Z_h,
\qquad
K_-\approx \phi\,Z_{\rm edge}\otimes Z_h,
\end{equation}
with $\phi=0.18$ and a numerically negligible generator-fit residual for $C_-$. The counter-rotating loop therefore realizes an Ising-like entangler in the extracted two-qubit frame rather than merely producing a large $\Dloc$ value. Along $C_-$ the quartet gap and pointwise quality remain open throughout the cycle, so the signal is not caused by a collapsing local frame.

\subsection{Momentum continuation and an experimental signature}

The fixed-$k_x$ ribbon isolates the helical mechanism. We now examine whether the same contrast persists when momentum is allowed to vary. To avoid bulk re-entry we continue the counter-rotating texture to the annulus
\begin{equation}
\mathcal A_{\rm BHZ}=\{(k_x,\vartheta):k_x\in[0.05,0.35],\ \vartheta\in[0,2\pi)\},
\qquad
(\theta_T,\theta_B)=(\vartheta,-\vartheta).
\end{equation}

The model is described by the same Hamiltonian \eqref{eq:Ham_BHZ}. Only the control manifold changes. The microscopic operators and the quartet extraction stay the same.

Across this annulus the quartet remains well resolved, with $\Delta_4^{\min}=0.31$, $\max\epsjoint=0.02$, and outer-two-row weight above $0.94$.

The loop dependence survives the momentum continuation. Over the full benchmark window,
\begin{equation}
\Dloc(C_+)\approx 0.01,
\qquad
\Dloc(C_-)\approx 0.4.
\end{equation}
The branch containing the representative counter-rotating loop therefore remains entangling while the co-rotating control stays almost local. The holonomy also remains close to the Ising form $e^{-i\phi Z_{\rm edge}\otimes Z_h}$: the best-fit distance stays below approximately $0.02$ across the annulus. The fixed-slice contrast is therefore not confined to a single high-symmetry slice.

The annulus also gives a simple output-state witness. Here $|++\rangle$ denotes the equal-superposition product state in the extracted local basis. Under $C_-$ this state acquires concurrence of approximately $0.4$, while the co-rotating control remains negligible over the same momentum window. The single-edge control remains intermediate, with concurrence of approximately $0.2$. This does not replace direct Wilson-loop or interferometric reconstruction of the full holonomy \cite{Sugawa2021,Tyner2024}. It shows instead that failure of local reduction can be converted into a two-qubit entanglement witness and leads naturally to the tomography protocol summarized in Section~\ref{sec:discussion}.

\begin{figure}[H]
  \centering
  \includegraphics[width=\textwidth]{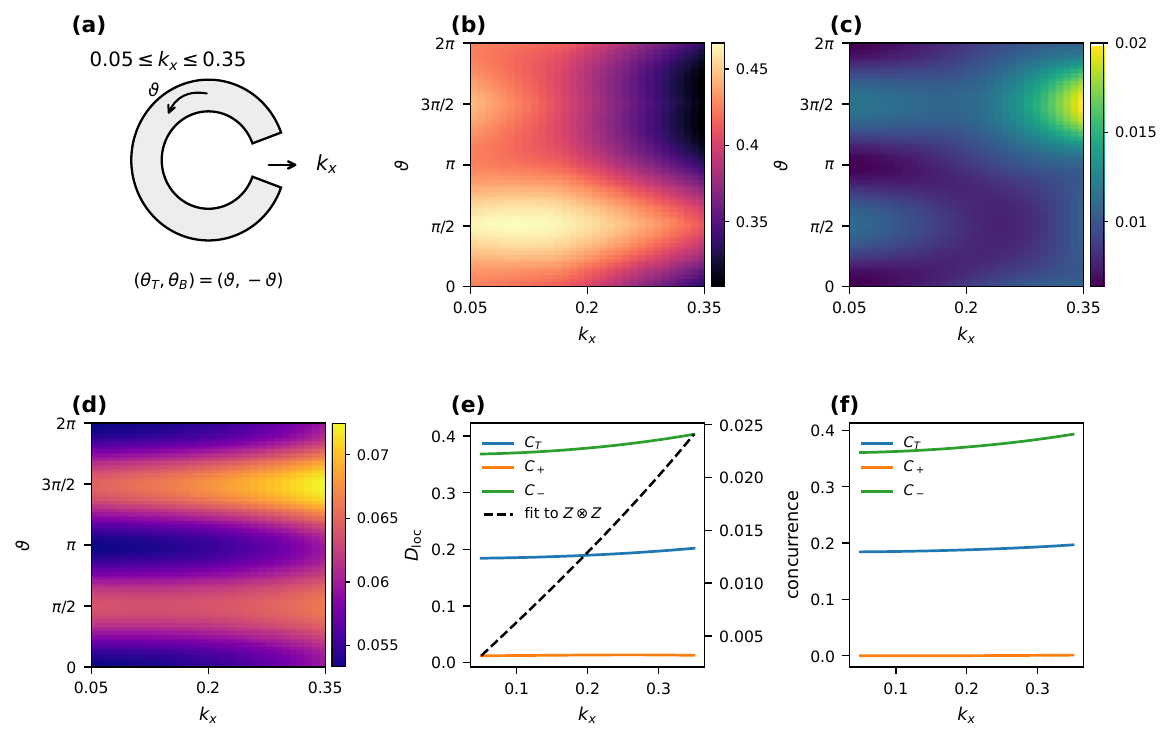}
  \caption{Momentum-resolved BHZ annulus. (a) Definition of the annulus $k_x\in[0.05,0.35]$ with counter-rotating texture $(\theta_T,\theta_B)=(\vartheta,-\vartheta)$; the curved and radial arrows indicate increasing $\vartheta$ and $k_x$, respectively. (b) Quartet gap. (c) Pointwise $\epsjoint$. (d) Angular component of the link-level distance to the local Lie algebra. (e) Fixed-$k_x$ loop distances for the single-edge, co-rotating, and counter-rotating controls; the dashed line is the local fit of $C_-$ to the Ising form $Z_{\rm edge}\otimes Z_h$, and $K$ again denotes the extracted helical pseudospin. (f) Concurrence generated from the product input $|++\rangle$ in the extracted local basis.}
  \label{fig:annulus}
\end{figure}

\FloatBarrier
\section{SSH chain: controlled-rotation benchmark on an edge quartet}\label{sec:ssh}

The SSH chain isolates the same mechanism in the numerically most stable setting of the three models. The same edge quartet again shows several reduction outcomes. A single-edge loop acts as a controlled rotation, while the anti-diagonal family gives the larger entangling response. Quartet isolation and split quality are strongest here, so the departure from the local subgroup can be followed without any collapse of the pointwise local frame.

The model is an open spinful dimerized chain of length $N=16$ with independent left and right edge angles,
\begin{align}
H_{\rm SSH} &= \sum_{n=1}^{N} \bigl(c_{nA}^\dagger T_1 c_{nB} + \text{h.c.}\bigr)
+ \sum_{n=1}^{N-1} \bigl(c_{nB}^\dagger T_2 c_{n+1,A} + \text{h.c.}\bigr) \\
&\quad + B\,c_{1A}^\dagger (\cos\theta_L\,\sigma_x+\sin\theta_L\,\sigma_y)c_{1A}
+ B\,c_{NB}^\dagger (\cos\theta_R\,\sigma_x+\sin\theta_R\,\sigma_y)c_{NB},
\end{align}
with
\begin{equation}
T_1=t_1 I + i\lambda_1\sigma_y,
\qquad
T_2=t_2 I + i\lambda_2\sigma_x.
\end{equation}
We use $(t_1,t_2)=(0.55,1.0)$, $(\lambda_1,\lambda_2)=(0.20,0.15)$, and $B=0.30$. The intended local structure is $(\text{edge})\otimes(\text{spin})$, extracted from compressed edge-position and microscopic spin observables.

Across the full torus, $\Delta_4^{\min}=0.26$, $\max\epsjoint=0.001$, and the edge weight stays above $0.60$. The loop dependence remains pronounced. For the symmetry-defined cycles
\begin{equation}
C_L:(\theta_L,\theta_R)=(t,0),\quad
C_R:(0,t),\quad
C_{\rm diag}:(t,t),\quad
C_{\rm anti}:(t,-t),
\qquad t\in[0,2\pi],
\end{equation}
we find
\begin{equation}
\Dloc(C_L)=\Dloc(C_R)=0.20,
\qquad
\Dloc(C_{\rm diag})=0.14.
\end{equation}
The anti-diagonal loop $C_{\rm anti}$ yields the largest response among these symmetry-defined cycles, with $\Dloc(C_{\rm anti})=0.38$ even though its maximal eigenphase matches that of $C_L$ to the precision shown here. This comparison is revisited in \Cref{tab:berry,tab:schmidt}. Among the single-edge loops, $C_L$ is used below as the representative controlled-rotation example, while supplementary scans place $C_L$, $C_{\rm anti}$, and the other named loops on nearby branches of the same torus. The same edge quartet therefore exhibits several reduction outcomes while the pointwise local frame remains well resolved across the torus. For the named loops, SSH shows the same ordering as BHZ: the single-edge cycle is intermediate, the anti-diagonal cycle is the larger entangler, and the canonical coordinates in Table~\ref{tab:cartan} show that both pairs lie on the same one-parameter edge of the Weyl chamber.

The representative single-edge loop admits a compact effective-generator description. Writing $P_L=(I+Z_{\rm edge})/2$ for the projector onto the left-edge block, its effective generator is well fit by
\begin{equation}
K_L\approx \phi\,P_L\otimes(\hat n\cdot\sigma),
\qquad
\phi=0.20,
\end{equation}
with $\hat n=(0.32,0.16,-0.93)$ and generator residual below $0.001$. The left block rotates while the spectator block remains nearly stationary, giving a controlled rotation on the extracted two-qubit space. Along $C_L$ the quartet gap stays above $0.26$, $\epsjoint<0.001$, and the nearest-neighbor frame overlap stays above $0.99$. The entangling response therefore survives stringent quality checks and varies weakly with system size over the range studied.

In this model, different closed loops act as local, controlled, or entangling operations on the extracted edge-spin quartet.

\begin{figure}[H]
  \centering
  \includegraphics[width=\textwidth]{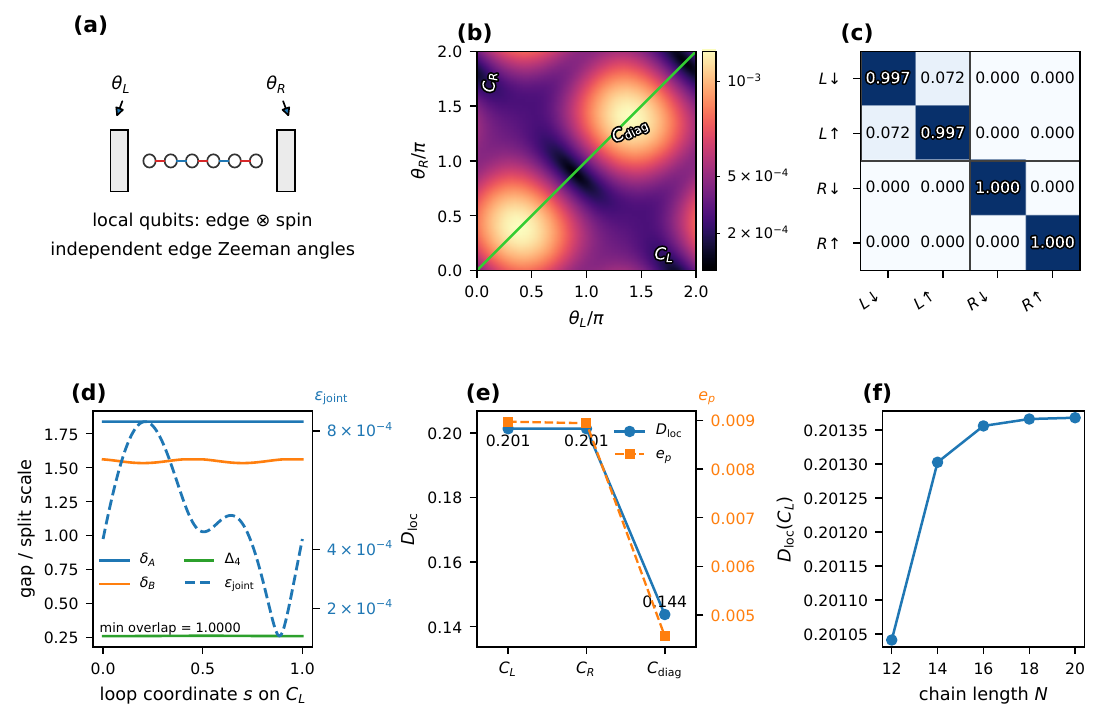}
  \caption{SSH benchmark. (a) Open spinful SSH chain with independent edge angles $(\theta_L,\theta_R)$. (b) $\epsjoint$ map on the edge-angle torus, with representative loops overlaid. (c) Left-loop holonomy in the extracted local basis. (d) Along-loop quality on $C_L$. (e) Representative loop comparison, shown as aligned $\Dloc$ and entangling-power $\ep$ axes. (f) Finite-size dependence of the single-edge response. Supplementary scans display nearby branches, including the anti-diagonal family.}
  \label{fig:ssh}
\end{figure}

\FloatBarrier
\section{BBH corner quartet: higher-order reduction failure}\label{sec:bbh}

BBH shows that the same question is not confined to first-order edge multiplets. The numerical margins are narrower than in SSH or BHZ, so this model is best read as a higher-order extension. Even so, the loop contrast is clear. Axis cycles remain almost local, while mixed cycles are entangling. In the shared two-qubit description, this is the first case with a finite second nonlocal coordinate.

\begin{figure}[h]
  \centering
  \includegraphics[width=0.97\textwidth]{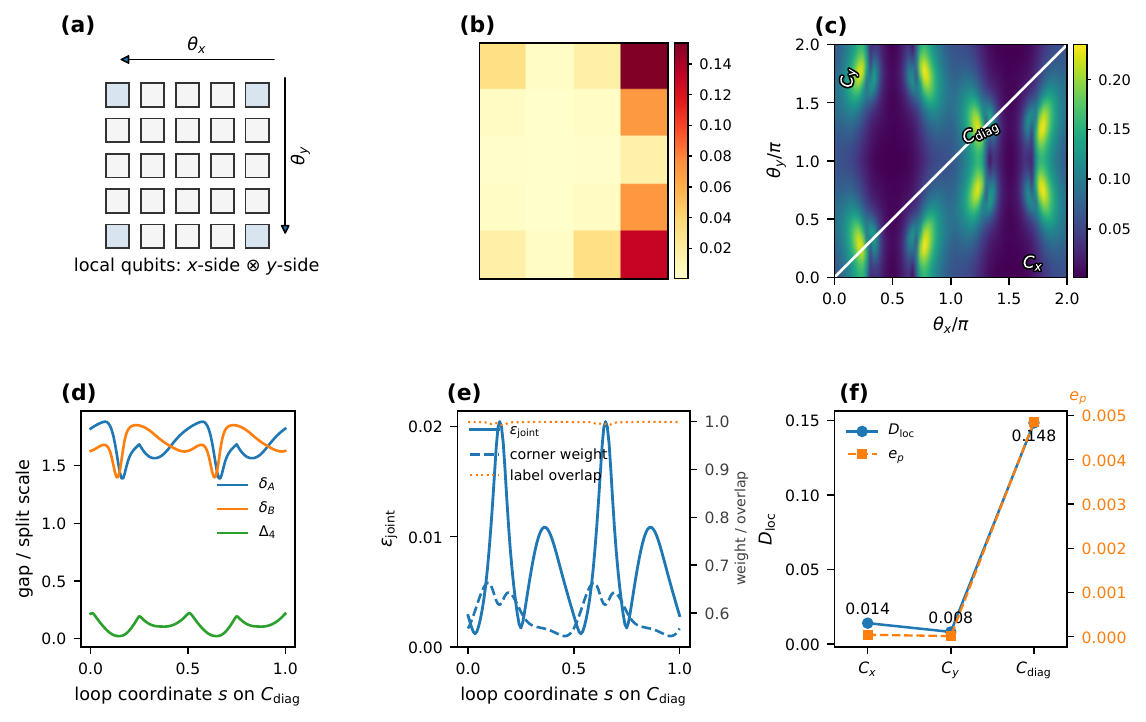}
  \caption{BBH benchmark. (a) Open BBH lattice with the corner-field torus $(\theta_x,\theta_y)$. (b) Representative corner-mode density at $(0,0)$; over the full torus the corner weight stays above $0.49$. (c) Map of the link-level distance to the local Lie algebra, with axis and diagonal loops overlaid. (d) Along the diagonal loop the quartet gap and split diagnostics remain open. (e) The same loop keeps $\epsjoint$ small while corner weight and label overlap remain high. (f) Loop comparison. The axis loops are almost local, while the diagonal family is entangling; the anti-diagonal family is symmetry related and agrees within the numerical resolution shown.}
  \label{fig:bbh}
\end{figure}

The higher-order extension is the open BBH quadrupole model on a $5\times5$ lattice (Fig.~\ref{fig:bbh}(a)) with $(\gamma_x,\gamma_y)=(0.3,0.3)$, $(\lambda_x,\lambda_y)=(1,1)$, and a corner-field torus generated by two angle-dependent corner masses of amplitudes $m_x=m_y=0.6$ \cite{BBH2017,Schindler2018}. The intended local structure is $(x\text{-side})\otimes(y\text{-side})$, extracted from coarse position observables compressed into the corner quartet.

With $\tau_0=\sigma_0=I_2$ and $c_{\mathbf r}$ the four component spinor on unit cell $\mathbf r=(x,y)$, the Hamiltonian is
\begin{align}
\begin{aligned}
H_{\rm BBH}(\theta_x,\theta_y)
&=
\sum_{\mathbf r} c_{\mathbf r}^\dagger
\bigl(\gamma_x\Gamma_4+\gamma_y\Gamma_2+\delta\Gamma_0\bigr)c_{\mathbf r}\\
&\quad +
\sum_{x=1}^{N_x-1}\sum_{y=1}^{N_y}
\bigl[c_{x,y}^\dagger \tfrac{\lambda_x}{2}(\Gamma_4-i\Gamma_3)c_{x+1,y}+\text{h.c.}\bigr]\\
&\quad +
\sum_{x=1}^{N_x}\sum_{y=1}^{N_y-1}
\bigl[c_{x,y}^\dagger \tfrac{\lambda_y}{2}(\Gamma_2-i\Gamma_1)c_{x,y+1}+\text{h.c.}\bigr]\\
&\quad +
\sum_{\mathbf r\in\mathrm{corners}}
\eta_{\mathbf r}\,
c_{\mathbf r}^\dagger
\Bigl[
m_x(\cos\theta_x M_1+\sin\theta_x M_2)
+
m_y(\cos\theta_y M_3+\sin\theta_y M_4)
\Bigr]
c_{\mathbf r},
\end{aligned}
\end{align}
where
\begin{align}
\Gamma_1&=-\tau_y\otimes\sigma_x,
&
\Gamma_2&=-\tau_y\otimes\sigma_y,
&
\Gamma_3&=-\tau_y\otimes\sigma_z,\\
\Gamma_4&=\tau_x\otimes\sigma_0,
&
\Gamma_0&=\tau_z\otimes\sigma_0,
&
M_1&=\tau_0\otimes\sigma_z,\\
M_2&=\tau_z\otimes\sigma_z,
&
M_3&=\tau_0\otimes\sigma_x,
&
M_4&=\tau_0\otimes\sigma_y.
\end{align}
Here $s_x(\mathbf r)=-1$ on the left boundary and $+1$ on the right boundary, while $s_y(\mathbf r)=-1$ on the bottom boundary and $+1$ on the top boundary. The corner sign is $\eta_{\mathbf r}=s_x(\mathbf r)s_y(\mathbf r)$. Both corner fields therefore carry the same corner sign $\eta_{\mathbf r}$ and act only on the four corner unit cells. The control torus rotates corner masses and does not modify the bulk hoppings.

This quartet is numerically less stable than in SSH or BHZ, so it serves as a higher-order extension rather than as a benchmark of comparable numerical quality. Even so, the regime of interest remains controlled: on the full torus we find $\Delta_4^{\min}=0.02$, $\max\epsjoint=0.02$, and corner weight above $0.49$. Along the diagonal entangling loop the quality improves further, with $\Delta_4>0.02$, $\epsjoint<0.02$, corner weight above $0.55$, and frame overlap above $0.99$.

For the symmetry-defined cycles
\begin{equation}
C_x:(\theta_x,\theta_y)=(t,0),\quad
C_y:(0,t),\quad
C_{\rm diag}:(t,t),\quad
C_{\rm anti}:(t,-t),
\qquad t\in[0,2\pi],
\end{equation}
the loop hierarchy is qualitatively distinct from SSH and BHZ. The axis loops remain nearly local,
\begin{equation}
\Dloc(C_x)=0.01,
\qquad
\Dloc(C_y)=0.008,
\end{equation}
whereas both mixed families are entangling,
\begin{equation}
\Dloc(C_{\rm diag})=\Dloc(C_{\rm anti})=0.15.
\end{equation}

Axis cycles mainly move one coarse side label at a time and stay near the local subgroup. Mixed cycles vary both coarse labels and couple all four corner sectors. In this model, reduction failure is therefore concentrated on mixed cycles rather than on axis motion. Consistently, the operator-Schmidt spectrum of $C_{\rm diag}$ requires four visible channels, unlike the nearly rank-two SSH and BHZ entanglers. The fitted generator is likewise spread across several tensor-product channels, as shown in the Supplementary analyses. In canonical two-qubit language, $C_{\rm diag}$ is not just larger than the axis loops. It is the first case in this set with a finite second nonlocal coordinate.

As a null-homotopic control, we also evaluated a center-based family of contractible circles,
\begin{equation}
\mathcal C_0(\rho):\quad
(\theta_1,\theta_2)/\pi=c_0+\rho(\cos t,\sin t),
\qquad c_0=(1,1),\qquad t\in[0,2\pi].
\end{equation}

The same center \(c_0=(1,1)\), radius list, and positive-orientation convention are used in BHZ, SSH, and BBH. Only the physical meaning of $(\theta_1,\theta_2)$ changes from model to model. The radius dependence is shown in the Supplementary analyses. In BBH, the contractible-disk scan develops a finite-radius branch with \(D_{\rm loc}=\mathcal O(10^{-1})\), while the corresponding BHZ and SSH circles remain much smaller on the same geometric family. Since these circles are null-homotopic, this response is not a fundamental-group effect. It comes from the nonlocal component of the quartet Berry curvature over the enclosed disk. Reversing the orientation gives the inverse holonomy, \(U_-\simeq U_+^\dagger\), and leaves \(\Dloc\) and \(\ep\) unchanged to numerical precision.

\FloatBarrier
\section{Why standard Berry data do not determine the obstruction\label{sec:Berry}}

Standard Berry data fail for a simple reason. Chern numbers, determinant phases, and eigenphase spectra summarize phases or spectra of the loop operator, but they do not ask whether that operator can be written as two independent single-qubit actions in the extracted basis. The obstruction studied here is therefore invisible to them in principle. The examples below show both kinds of mismatch. Nearly identical reduced Berry data can hide different gate classes, and large reduced phases can occur on loops that remain almost local.

The same point holds for any factorization type $\mathbf d$. Spectral data classify a unitary only up to conjugation in $\U(D)$. Locality is a statement about membership in the much smaller product subgroup $G_{\mathbf d}$. Two holonomies can therefore be isospectral, or can have the same determinant phase, while belonging to different product-subgroup classes. The two-qubit cases below are explicit low-dimensional examples of this mismatch.

This point should be separated from the usual use of Berry curvature.  The quartet curvature $F_A=dA+A\wedge A$ is a standard gauge-theoretic object, and its integral controls small contractible holonomies.  What is specific to the present analysis is the subgroup-relative question: whether the resulting holonomy, or the curvature component that generates it to leading order, lies inside the local algebra $\mathfrak g_{\rm loc}$.  The obstruction is therefore not simply ``nonzero Berry curvature''. It is nonzero curvature or monodromy with a component transverse to the extracted local subgroup.

The first indication appears at the torus level: using the Fukui--Hatsugai--Suzuki discretization \cite{Fukui2005}, the first Chern number of each quartet bundle is numerically zero, and so are the Chern numbers of the natural rank-two blocks extracted from the local frame. In the present data these quantities are $0$ within numerical resolution. These are not the bulk topological indices of the underlying lattice models; they are Chern numbers of the low-energy quartet bundle over the control torus. For the present question they are silent.

The same is true at the loop level for Abelian phases. For all representative loops in \Cref{tab:berry}, the determinant phase $\arg\det U(\mathcal C)$ is numerically zero. Nevertheless $\Dloc$ remains finite for the physically important cycles. The local-subgroup obstruction therefore cannot be reduced to a $\mathrm U(1)$ Berry phase.

Wilson-loop eigenphases are more informative than a scalar phase, but they still do not determine whether transport is local or entangling. In SSH, $C_L$ and $C_{\rm anti}$ have nearly identical maximal eigenphase, approximately $0.20$, even though their $\Dloc$ values differ by almost a factor of two. In BBH, the almost local loop $C_y$ carries large eigenphases $\pm0.83$ despite $\Dloc(C_y)=0.008$. BHZ provides the same comparison in a transparent form. At the fitted low-energy level, the co-rotating and counter-rotating generators $I\otimes Z_h$ and $Z_{\rm edge}\otimes Z_h$ both produce the eigenphase multiset $\{\pm\phi,\pm\phi\}$; the equality is exact because the two generators differ in subgroup structure, not in eigenvalue content. The microscopic loops inherit that near-degeneracy: to the precision shown in \Cref{tab:cartan}, $C_+$ and $C_-$ have the same sorted eigenphase quadruplet, even though $\Dloc$ changes from $0.01$ to $0.37$. Canonical Cartan coordinates then separate the loop families: SSH and BHZ lie on the same one-parameter edge of the Weyl chamber, whereas BBH $C_{\rm diag}$ develops a finite second nonlocal coordinate. Supplementary operator-Schmidt spectra lead to the same hierarchy without relying only on the fitted-generator picture or on the Frobenius distance.

\begin{table}[t]
\centering
\caption{\textbf{Standard Berry data versus the local-reduction obstruction.} Entangling powers are exact two-qubit values obtained from the canonical coordinates of \Cref{tab:cartan}; independent Monte Carlo checks agree within numerical resolution.}
\label{tab:berry}
\small
\setlength{\tabcolsep}{4pt}
\begin{tabular}{lccc}
\toprule
bundle & quartet $\mathrm{Ch}_1$ & block A $\mathrm{Ch}_1$ & block B $\mathrm{Ch}_1$ \\
\midrule
BHZ quartet / blocks & $\approx 0$ & $\approx 0$ & $\approx 0$ \\
SSH quartet / blocks & $\approx 0$ & $\approx 0$ & $\approx 0$ \\
BBH quartet / blocks & $\approx 0$ & $\approx 0$ & $\approx 0$ \\
\bottomrule
\end{tabular}

\vspace{2mm}
\begin{tabular}{lccccc}
\toprule
model & loop & $\arg\det U(\mathcal C)$ & $\max_j|\phi_j|$ & $\Dloc(\mathcal C)$ & $\ep(\mathcal C)$ \\
\midrule
BHZ & $C_T$ & $\approx 0$ & 0.18 & 0.18 & 0.007 \\
BHZ & $C_+$ & $\approx 0$ & 0.18 & 0.01 & $<0.001$ \\
BHZ & $C_-$ & $\approx 0$ & 0.18 & 0.37 & 0.03 \\
SSH & $C_L$ & $\approx 0$ & 0.20 & 0.20 & 0.009 \\
SSH & $C_{\rm anti}$ & $\approx 0$ & 0.20 & 0.38 & 0.03 \\
BBH & $C_y$ & $\approx 0$ & 0.83 & 0.008 & $<0.001$ \\
BBH & $C_{\rm diag}$ & $\approx 0$ & 1.26 & 0.15 & 0.005 \\
\bottomrule
\end{tabular}
\normalsize
\end{table}

\FloatBarrier
\section{Conclusion and experimental outlook}\label{sec:discussion}

Across the three models, a pointwise local factorization does not fix the gate class of the loop holonomy. Changing only the loop moves the same quartet between almost local motion, controlled rotations concentrated on one block, and more distributed entanglers. These cases cannot be ordered by a single scalar quantity. Once reduction fails, canonical coordinates place SSH and BHZ on the same one-parameter edge of the two-qubit Weyl chamber, while the BBH mixed cycle develops a finite second nonlocal coordinate.

The quartet language is the smallest explicit form of a dimension-independent statement. For a higher-dimensional boundary multiplet with pointwise labels $(d_1,\ldots,d_r)$, the same analysis replaces $\U(2)\otimes\U(2)$ by the embedded product subgroup $G_{\mathbf d}$ and asks whether the holonomy reduces to that subgroup. The same issue should arise whenever the connection of the isolated multiplet has curvature or monodromy with a component transverse to $\mathfrak g_{\mathbf d}$.

The analysis has two layers. $\Dloc$ asks whether the microscopic holonomy itself lies in the local subgroup. Cartan coordinates, operator-Schmidt spectra, and fitted generators then describe the gate geometry that remains after the holonomy has left that subgroup. Chern numbers, determinant phases, and eigenphase spectra do not answer the subgroup-membership question on their own.

This loop dependence is the microscopic form of the gluing problem studied in Ref.~\cite{Ikeda2026}. The mathematical formulation is given in the Supplementary analyses. For the present paper, the main point is that a pointwise local factorization can remain well defined while the closed-loop holonomy fails to remain local.

A few limits of the analysis should be stated plainly. The calculations remain quartet calculations. The higher-dimensional discussion extends the same diagnostic framework but does not add a separate numerical study. Quartet selection is guided by symmetry and checked a posteriori through localization and split diagnostics. The BHZ annulus covers a benchmark momentum window rather than the full Brillouin zone. BBH is physically informative but numerically less stable than SSH or BHZ. None of these points changes the qualitative conclusion. The trivial-side SSH and BHZ controls, and the non-HOTI BBH control, all lose the localization and split-quality conditions before any loop-level local-subgroup interpretation becomes reliable. In that sense, the present analysis relies on topological boundary quartets.

Possible extensions include larger momentum windows, more realistic helical-edge models on the noninteracting side, higher-dimensional boundary multiplets and many-body edge or ladder quartets obtained from exact diagonalization or tensor-network calculations. The present loop contrast also admits an experimental formulation. Recent experiments already demonstrate non-Abelian adiabatic geometric control and Wilson-loop readout in synthetic settings \cite{Leroux2018,Sugawa2021}, and related Wilson-loop measurement protocols have been formulated for crystalline systems \cite{Tyner2024}.

\paragraph{Proposed experimental implementation.}

A practical measurement can be organized in three stages. One first identifies a control window in which the quartet remains isolated by monitoring the quartet gap $\Delta_4$, the split diagnostics, and the relevant localization weight, namely the outer-edge weight in BHZ, the edge weight in SSH, and the corner weight in BBH. Over that window, the compressed observables $(O_A,O_B)$ define the extracted local basis and therefore the product reference states of the effective two-qubit description. One then prepares the product input $|++\rangle$ in that basis, applies a chosen closed loop, and performs standard two-qubit output-state tomography \cite{James2001}. Operationally, this uses basis rotations within the quartet together with readout of the compressed observables that define the two labels. Reconstructing $\rho_{\rm out}(\mathcal C)$ gives a direct witness of local-reduction failure through the output concurrence. In BHZ the simplest comparison is $C_+$ versus $C_-$. In SSH it is $C_L$ versus $C_{\rm anti}$. In BBH it is an axis loop versus $C_{\rm diag}$.

For a direct determination of the holonomy itself, the witness stage can be upgraded to full quantum process tomography \cite{ChuangNielsen1997} on the extracted two-qubit manifold. Preparing the tensor-product inputs $\{|0\rangle,|1\rangle,|+\rangle,|+i\rangle\}^{\otimes 2}$ and performing the same output tomography reconstructs the effective loop channel, from which one can recover the nearest unitary representative, $\Dloc$, and the Cartan coordinates. In a microscopic adiabatic realization, any dynamical contribution can in principle be removed by comparison with a reference sequence or by interferometric calibration of the holonomy \cite{Sugawa2021,Tyner2024}, so that the reconstructed operator isolates the geometric holonomy rather than the total phase accumulation.

For a higher-dimensional implementation, the reconstruction protocol is unchanged in form. One prepares a tomographically complete set of product inputs in $\bigotimes_\alpha\mathbb C^{d_\alpha}$, reconstructs the loop channel on the isolated multiplet, and computes $D_{\mathbf d}$ relative to $G_{\mathbf d}$. The witness stage would then use partition-dependent entanglement witnesses or operator-Schmidt diagnostics in place of two-qubit concurrence, while the full process stage would test subgroup membership of the reconstructed holonomy directly.

The three models are not equally demanding experimentally. SSH appears to be the simplest target in the present data because quartet isolation and edge-spin readout margins are strongest there. BHZ requires independent control of top and bottom edge textures and readout calibrated to the extracted helical pseudospin. BBH requires corner-resolved preparation and detection and is therefore more demanding. The extracted two-qubit holonomies can also be implemented on programmable quantum hardware. The accompanying Qiskit implementation realizes representative BHZ, SSH, and BBH extracted-loop circuits, together with output-state tomography and, when needed, full two-qubit process tomography. The same circuit definitions can be run on local Aer simulation or on IBM Quantum backends by changing only the backend specification, with common classical post-processing. In all these forms, the main result is the same. A single boundary quartet can support distinct reduction outcomes under different closed loops.

\section*{Data and code availability}
Source data and code for the figure panels, benchmark tables, canonical-coordinate summaries, robustness scans, and quantum circuit implementation with Qiskit are included in the accompanying source bundles in authors’ GitHub repository:\\
\url{https://github.com/IKEDAKAZUKI/Entanglement-Obstruction-in-Condensed-Matter}.

\section*{Acknowledgments}
This work was supported by the NSF under Grant No. OSI-2328774 (KI), by the Israeli Science Foundation Excellence Center, the US-Israel Binational Science Foundation, the Israel Ministry of Science (YO).

%\clearpage
\appendix
\section{Methods}

\subsection*{Quartet projection and pointwise local frame}

The construction used in the main text is the rank-four specialization of a general rank-$D$ projection procedure.  Given an isolated multiplet with $D=\prod_\alpha d_\alpha$, one forms $P_D(\lambda)$ from the selected states and compresses a fixed family of physical observables whose joint spectral structure resolves the desired subsystem labels.  A viable higher-dimensional local frame requires finite spectral splits for these labels, small incompatibility among the compressed observables, and stable localization of the selected multiplet.  The quartet benchmarks below set $D=4$ and $\mathbf d=(2,2)$, for which two observables are sufficient to resolve the ordered product basis used in the numerical analysis.

For all three models we work with a smooth Hamiltonian family $H(\lambda)$ and an isolated rank-four projector built from the four low-energy states nearest zero energy,
\begin{equation}
P_4(\lambda)=\sum_{a=1}^4 |u_a(\lambda)\rangle\langle u_a(\lambda)|.
\end{equation}
We do not regard the resulting local frame as a unique microscopic tensor product. Instead, we ask whether the loop holonomy stays inside the local subgroup extracted from two physically motivated observables $O_A$ and $O_B$ whose quartet compressions exhibit a well-resolved $2+2$ split. The observables themselves need not be binary in the full Hilbert space: in BHZ we use position $y$ and spin $s_z$, and in BBH we use coarse position observables. Binary labels emerge only after compression to the isolated quartet. The pair $(O_A,O_B)$ is fixed once per model from locality and symmetry considerations before any loop-level analysis and is not tuned loop by loop or path by path to maximize $\Dloc$.

The compressed observables are
\begin{equation}
\widetilde O_A=P_4 O_A P_4,
\qquad
\widetilde O_B=P_4 O_B P_4,
\end{equation}
and are sequentially diagonalized to define a pointwise local frame $F(\lambda)\in\mathbb C^{N\times4}$. In SSH these observables resolve edge and spin, in BBH they resolve coarse $x$- and $y$-side labels, and in BHZ they resolve ribbon side and the spin-derived helical pseudospin described in the main text. The construction uses a fixed order: $O_A$ defines the coarse two-block split and $O_B$ resolves each block internally. We do not claim uniqueness of this extraction; rather, the qualitative loop hierarchy is stable across the smooth observable deformations reported in Supplementary Fig.~\ref{fig:robust}. We also checked the explicit order swap $O_A\leftrightarrow O_B$: in SSH and BHZ the hierarchy among almost local, intermediate, and entangling loops survives with only small numerical shifts, and in BBH the axis-versus-mixed distinction survives as well (\Cref{tab:orderswap}).

The pointwise quality of this factorization is monitored by the quartet gap $\Delta_4$ and by the normalized commutator
\begin{equation}
\epsjoint(\lambda)=\frac{\norm{[\widetilde O_A,\widetilde O_B]}_F}{\mathrm{range}(\widetilde O_A)\,\mathrm{range}(\widetilde O_B)}.
\end{equation}
Here $\norm{X}_F=\sqrt{\Tr(X^\dagger X)}$ is the Frobenius norm and $\mathrm{range}(X)=\lambda_{\max}(X)-\lambda_{\min}(X)$ is evaluated inside the quartet. Small $\epsjoint$, combined with a finite quartet gap, indicates that the local two-qubit description is well defined. After diagonalizing $\widetilde O_A$ we use its middle gap to split the quartet into two two-dimensional blocks. Writing the ordered eigenvalues of $\widetilde O_A$ as $a_1\le a_2\le a_3\le a_4$, the corresponding split diagnostic is
\begin{equation}
\delta_A=a_3-a_2.
\end{equation}
Inside each $2\times2$ block we diagonalize $\widetilde O_B$ and define the block-wise split diagnostic $\delta_B$ as the smaller of the two level splittings. These quantities appear in the along-loop quality panels.

\subsection*{Loop holonomies and local-subgroup diagnostics}

For neighboring parameter points we define the unitary link
\begin{equation}
\mathcal U_{\lambda\to\lambda'}=\mathrm{polar}\bigl(F(\lambda)^\dagger F(\lambda')\bigr),
\end{equation}
whose ordered product around a closed discretized loop gives the extracted Wilczek--Zee holonomy $U(\mathcal C)$. The main loop diagnostic is the distance to the local subgroup,
\begin{equation}
\Dloc(\mathcal C)=\min_{A,B\in\U(2)}\norm{U(\mathcal C)-A\otimes B}_F.
\end{equation}

For a general factorization type, the numerical objective becomes
\begin{equation}
D_{\mathbf d}(\mathcal C)=\min_{V_\alpha\in\U(d_\alpha)}\norm{U(\mathcal C)-V_1\otimes\cdots\otimes V_r}_F,
\end{equation}
with optional minimization over physically allowed permutations of identical factors.  In the two-qubit benchmarks this reduces to the displayed $\Dloc$ and, as checked explicitly, allowing the qubit swap does not change the reported classification.

We use entangling power $\ep$ \cite{Zanardi2000} only as a secondary descriptor. For the representative named loops quoted in the main text and tables, $\ep$ is evaluated exactly from the canonical Cartan coordinates $(c_1,c_2,c_3)$ after removing global phase,
\begin{equation}
\ep=\frac{1}{18}\Bigl[3-\cos(2c_1)\cos(2c_2)-\cos(2c_2)\cos(2c_3)-\cos(2c_3)\cos(2c_1)\Bigr],
\end{equation}
which is the standard two-qubit linear-entropy entangling power in the Cartan-coordinate convention used in \Cref{tab:cartan} \cite{Zanardi2000,Makhlin2002,Zhang2003}. Independent Monte Carlo checks agree with these exact values within numerical resolution. For all representative loops reported in the main text, the same minimum is obtained even if the minimization is enlarged by allowing an explicit qubit swap; the observed nonlocality is therefore not a qubit-relabeling artifact.

The link-level nonlocal content plotted in some torus and annulus maps is built from the discrete Berry connection on a directed grid step $\delta\lambda$,
\begin{equation}
A(\lambda;\delta\lambda)=\frac{1}{|\delta\lambda|}\,\log\!\bigl[\mathcal U_{\lambda\to\lambda+\delta\lambda}\bigr],
\end{equation}
where the principal anti-Hermitian logarithm is used. We compare $A$ with the local Lie algebra
\begin{equation}
\mathfrak g_{\rm loc}=\{X_A\otimes I + I\otimes X_B + i\alpha I:\,X_A,X_B\in\mathfrak{su}(2),\ \alpha\in\mathbb R\},
\end{equation}
and write $\Pi_{\rm loc}$ for the Frobenius-type orthogonal projector onto $\mathfrak g_{\rm loc}$, implemented by orthogonalizing and normalizing the anti-Hermitian set $\{iI,\,i\sigma_\mu\otimes I,\,iI\otimes\sigma_\mu\}$ with respect to the Frobenius inner product.

In higher dimensions this algebra is replaced by $\mathfrak g_{\mathbf d}$, generated by anti-Hermitian operators acting on one factor at a time together with the global phase.  The projector $\Pi_{\rm loc}$ is then replaced by the Frobenius-orthogonal projector $\Pi_{\mathbf d}$ onto that algebra.

The plotted quantity is
\begin{equation}
\Nloc(\lambda)=\norm{A(\lambda;\delta\lambda)-\Pi_{\rm loc}A(\lambda;\delta\lambda)}_F,
\end{equation}
averaged over the coordinate directions of the underlying torus or annulus grid. The main conclusions do not depend on quantitative use of $\Nloc$, but the maps indicate where nonlocal transport is concentrated in parameter space.

Whenever we quote an effective generator fit, we use the Hermitian principal generator $K$ defined from the spectral logarithm of $U(\mathcal C)$. All representative loops satisfy $\max_j|\phi_j|<1.3<\pi$, so the branch is unambiguous on the reported cycles. The quoted generator-fit distance is the Frobenius residual $\norm{K-K_{\rm fit}}_F$, while the complementary operator-Schmidt spectra are obtained by reshaping the holonomy as a two-qubit operator and taking its Schmidt singular values.

\subsection*{Numerical conventions and quartet selection}

All torus maps use endpoint-free periodic grids, so the $0$ and $2\pi$ seam is not drawn twice. Loop holonomies are computed on closed grids including the endpoint. The representative named loops are the symmetry-defined axis, diagonal, or counter-diagonal cycles on the corresponding control torus; they are not selected a posteriori to maximize $\Dloc$. The supplementary continuous-$\eta$ scan is included only to indicate how these reference loops extend into nearby branches.

The nearest-neighbor frame-overlap diagnostic used in the along-loop quality panels is
\begin{equation}
\mathrm{ov}_{j,j+1}=\min_{a=1,\dots,4}\bigl|\langle f_a(\lambda_j)|f_a(\lambda_{j+1})\rangle\bigr|,
\end{equation}
after consistent ordering and phase fixing of the extracted basis. In the BBH quality panel the same quantity is labeled ``label overlap'' to fit the available figure space. The localization weights are model specific: SSH edge weight is the average quartet weight on the two terminal edge orbitals, BHZ outer-two-row weight is the average quartet weight on the first and last two lattice rows, and BBH corner weight is the average quartet weight on the four corner unit cells. The pointwise frame is phase-fixed only after the local basis has been extracted, so the local-subgroup diagnostics remain basis-covariant inside the quartet.

The quartet-selection rule is intentionally conservative. At each parameter point we choose the four states closest to zero energy, then verify that the resulting quartet has the intended edge or corner localization and that the compressed observables exhibit finite splitting. This is sufficient for the benchmark regimes studied here. For BHZ, the additional restriction to a finite momentum window around $k_x=0$ keeps the helical quartet spectrally separated over both the edge-angle torus and the momentum-angle annulus while avoiding bulk re-entry outside the benchmark window.

\FloatBarrier
\section{Supplementary analyses}

\subsection*{Effective generators and continuous loop families}

\begin{figure}[H]
  \centering
  \includegraphics[width=\textwidth]{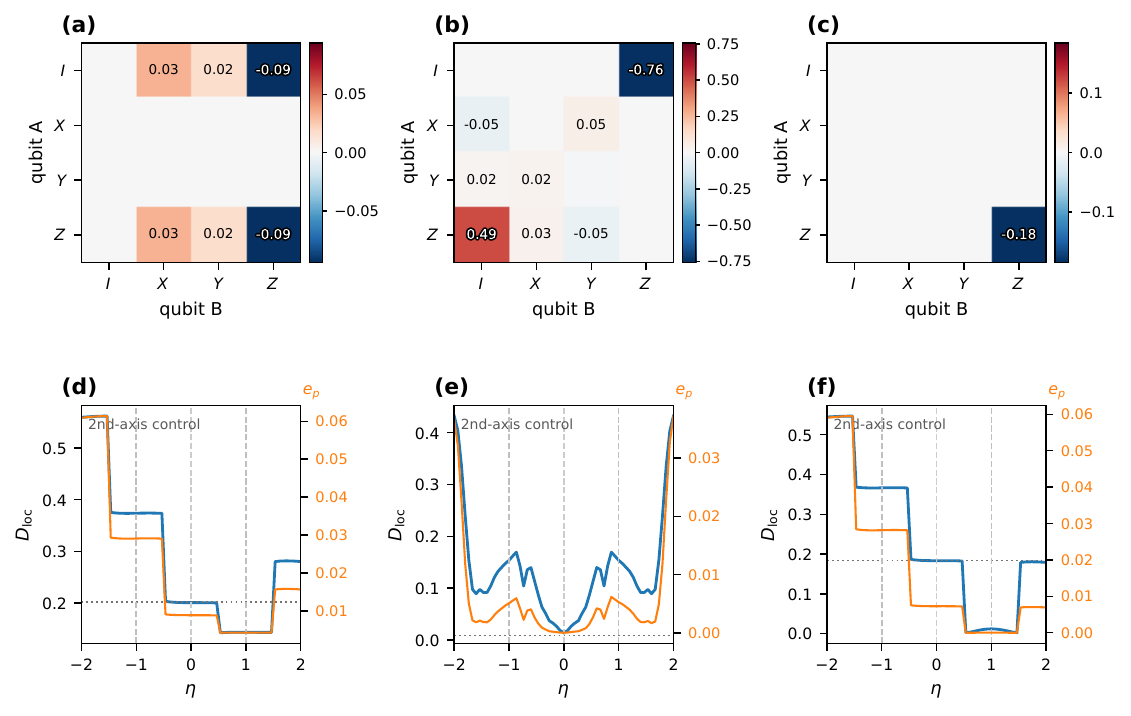}
  \caption{Mechanisms and continuous loop families. (a) Effective generator coefficients for the SSH left loop, dominated by a controlled-rotation form. (b) Effective generator coefficients for the BBH diagonal loop, with weight distributed across several higher-order channels. (c) Effective generator coefficients for the BHZ counter-rotating loop, dominated by the edge-helical Ising channel $Z_{\rm edge}\otimes Z_h$; the panel notation $Z_K$ denotes the same helical-pseudospin factor. (d--f) Continuous scans of $(\theta_1,\theta_2)=(t,\eta t)$ over $\eta\in[-2,2]$. For non-integer $\eta$, the path is closed by a short return segment. The blue curve shows $\Dloc$, the orange curve shows entangling power $\ep$, dotted horizontal lines denote the second-axis control, and dashed vertical lines mark $\eta=-1,0,1$. These scans show connected branches in slope space rather than an optimization over all closed loops at fixed slope.}
  \label{fig:mechloops}
\end{figure}

The effective generators clarify the physical interpretation of the benchmark loops. For SSH, the left-loop generator is concentrated in the $P_L\otimes(\hat n\cdot\sigma)$ sector, consistent with a controlled-rotation picture. For BBH, the diagonal-loop generator is distributed across several tensor-product channels rather than being dominated by a single one, which is why the response concentrates on mixed cycles. For BHZ, the representative counter-rotating loop isolates the edge-helical Ising term with a small residual.

A complementary basis-independent view is given by the operator-Schmidt spectra in \Cref{tab:schmidt}. The co-rotating BHZ loop is nearly rank one, as expected for an almost local gate. The BHZ counter-rotating loop and the SSH single-edge loop are both close to rank two, consistent with Ising-like and controlled-rotation structures, respectively, while the SSH anti-diagonal loop is a larger rank-two entangler. By contrast, the BBH diagonal loop requires four visible Schmidt channels, reinforcing the picture of a higher-order response spread across several channels. Operator-Schmidt data are not a complete local-equivalence classification of two-qubit gates \cite{Makhlin2002,Zhang2003}; here they are used as an additional basis-independent check rather than as a replacement for full canonical invariants.

\begin{table}[H]
\centering
\caption{\textbf{Operator-Schmidt spectra of representative holonomies.} Singular values $s_1\ge s_2\ge s_3\ge s_4$ of the $4\times4$ holonomies reshaped as two-qubit operators. The spectra provide a comparison of nonlocal structure that does not rely on the fitted generators.}
\label{tab:schmidt}
\small
\setlength{\tabcolsep}{5pt}
\begin{tabular}{lccccc}
\toprule
model & loop & $s_1$ & $s_2$ & $s_3$ & $s_4$ \\
\midrule
BHZ & $C_+$ & 2.00 & 0.01 & $<0.001$ & $<0.001$ \\
BHZ & $C_-$ & 1.97 & 0.37 & $<0.001$ & $<0.001$ \\
SSH & $C_L$ & 1.99 & 0.20 & $<0.001$ & $<0.001$ \\
SSH & $C_{\rm anti}$ & 1.97 & 0.37 & $<0.001$ & $<0.001$ \\
BBH & $C_y$ & 2.00 & 0.008 & 0.002 & $<0.001$ \\
BBH & $C_{\rm diag}$ & 2.00 & 0.14 & 0.06 & 0.004 \\
\bottomrule
\end{tabular}
\normalsize
\end{table}

To place the results in standard two-qubit language, \Cref{tab:cartan} lists canonical Cartan/Weyl-chamber coordinates after factoring out global phase \cite{Makhlin2002,Zhang2003}. Combined with the exact entangling powers reported in \Cref{tab:berry}, these coordinates reveal a common pattern: $C_T$ and $C_L$ occupy approximately the same one-parameter edge of the chamber, $C_-$ and $C_{\rm anti}$ move farther along that same edge, and $C_{\rm diag}$ is the first case here with a finite second Cartan coordinate. In that standard local-equivalence language, the loop hierarchy remains separated without privileging $\Dloc$, $\ep$, or the fitted generators.

\begin{table}[H]
\centering
\caption{\textbf{Canonical two-qubit coordinates of representative holonomies.} Sorted eigenphases are quoted in radians at a precision sufficient for the comparisons used in the text. The Cartan coordinates $(c_1,c_2,c_3)$ are given after removing global phase and mapping the nonlocal part to the Weyl chamber $0\le c_3\le c_2\le c_1\le \pi/2$ \cite{Makhlin2002,Zhang2003}.}
\label{tab:cartan}
\small
\setlength{\tabcolsep}{4pt}
\begin{tabular}{llp{5.8cm}c}
\toprule
model & loop & sorted eigenphases & Cartan $(c_1,c_2,c_3)$ \\
\midrule
BHZ & $C_+$ & $\{-0.18,-0.18,0.18,0.18\}$ & $(0.01,0,0)$ \\
BHZ & $C_T$ & $\{-0.18,0,0,0.18\}$ & $(0.18,<0.01,0)$ \\
BHZ & $C_-$ & $\{-0.18,-0.18,0.18,0.18\}$ & $(0.37,0,0)$ \\
SSH & $C_L$ & $\{-0.20,0,0,0.20\}$ & $(0.20,<0.01,0)$ \\
SSH & $C_{\rm anti}$ & $\{-0.20,-0.20,0.20,0.20\}$ & $(0.38,<0.01,0)$ \\
BBH & $C_y$ & $\{-0.83,-0.83,0.83,0.83\}$ & $(0.008,0.002,0)$ \\
BBH & $C_{\rm diag}$ & $\{-1.26,-0.27,0.27,1.26\}$ & $(0.14,0.06,0)$ \\
\bottomrule
\end{tabular}
\normalsize
\end{table}

The bottom row of \Cref{fig:mechloops} complements the named cycles by scanning the one-parameter family
\begin{equation}
(\theta_1,\theta_2)=(t,\eta t),
\qquad
\eta\in[-2,2].
\end{equation}
For integer $\eta$ this is a closed straight-slope loop on the torus. For non-integer $\eta$, however, the path is closed only after appending a short return segment. The continuous scan is therefore an auxiliary visualization of connected branches in slope space, not a length-normalized optimization over all closed loops at fixed slope. The main-text claims are tied to the symmetry-defined representative loops, while the auxiliary scan shows how those representative loops sit inside nearby branches.

For SSH, the anti-diagonal branch around $\eta\approx -1$ attains larger values than the diagonal branch near $\eta\approx +1$, while the axis controls remain between them. For BBH, the axis sector stays near local while mixed cycles carry the larger response over a broad region of slope space. For BHZ, the separation is strongest: the branch containing the representative counter-rotating loop near $\eta=-1$ remains entangling, the co-rotating branch near $\eta=+1$ stays almost local, and the single-edge control remains between those two branches. Values outside the exact symmetry points should therefore be read as branch diagnostics rather than as a ranking of closed loops by a universal cost function.

\subsection*{Contractible center-based controls}

\begin{figure}[H]
  \centering
  \begingroup
  \includegraphics[width=\textwidth]{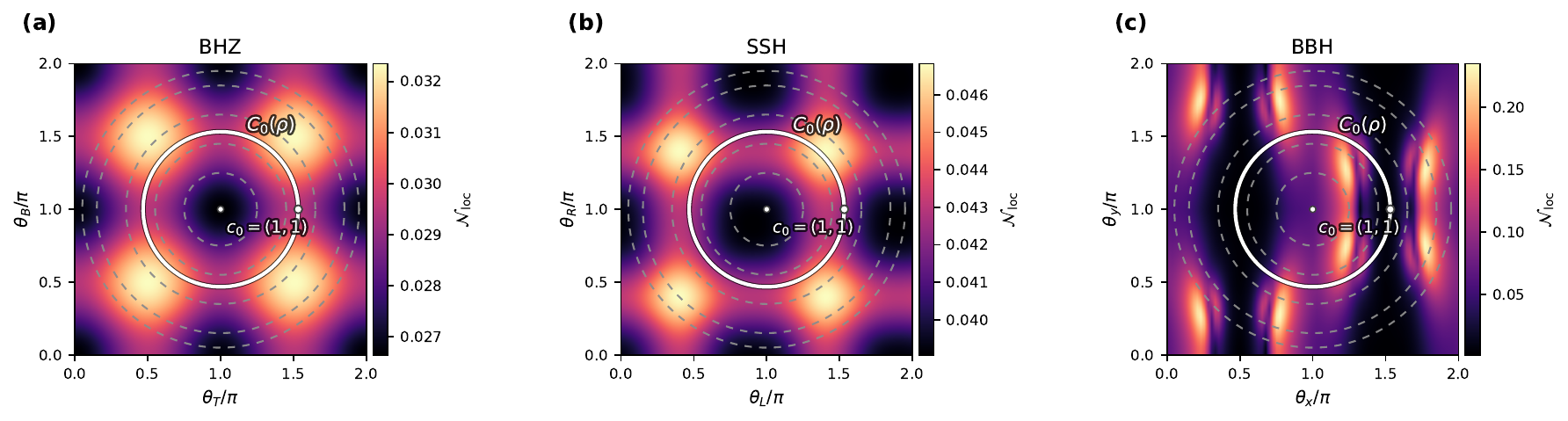}
  \caption{\textbf{Contractible center-based family on the control tori.}  The same center \(c_0=(1,1)\) and radius scan \(\mathcal C_0(\rho)\) are used in BHZ, SSH and BBH.  Dashed curves show the scanned radii; the solid curve marks the benchmark circle \(\rho=0.53\) used for the orientation check.  The heat maps show refined link-level distances \(\mathcal N_{\rm loc}\), plotted with periodic edge filling to avoid artificial blank boundaries.}
  \label{fig:contractibleloops}
  \endgroup
\end{figure}

\begin{figure}[H]
  \centering
  \begingroup
  \includegraphics[width=0.76\textwidth]{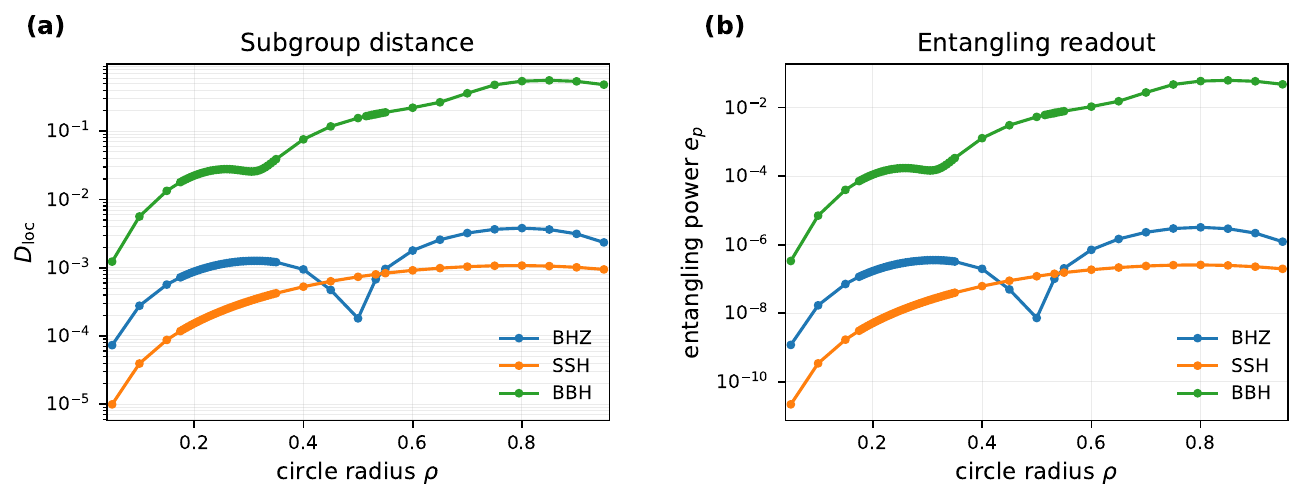}
  \caption{\textbf{Radius dependence of contractible-loop nonlocality.}  The plotted values use the positive orientation.  The reverse orientation gives the inverse holonomy and the same \(\Dloc\) and \(e_p\) to numerical precision.}
  \label{fig:contractibleradius}
  \endgroup
\end{figure}

To separate loop homotopy from Berry-curvature effects without selecting a single contractible loop by hand, we computed a center-based radius scan in each control torus.  In normalized coordinates the family is
\begin{equation}
\mathcal C_0(\rho):\quad
(\theta_1,\theta_2)/\pi=(1,1)+\rho(\cos t,\sin t),
\qquad \rho=0.05,0.10,\ldots,0.95.
\end{equation}
All three models use the same center, the same radius scan and the same positive-orientation convention.  \Cref{fig:contractibleloops} overlays this common family on \(\mathcal N_{\rm loc}\) maps with periodic edge filling, and \Cref{fig:contractibleradius,tab:contractible} gives the radius dependence and an orientation check at the representative radius \(\rho=0.53\).  The result is not isolated to a single tuned circle: BBH develops an extended finite-radius branch with \(D_{\rm loc}=\mathcal O(10^{-1})\) and continues to larger values at larger radii, while BHZ and SSH remain small on the same geometric family.

Here the word curvature refers only to the non-Abelian Berry curvature of the quartet connection.  The control torus is a flat parameter manifold; nevertheless the rank-four eigenstate bundle over it can carry a non-flat connection.  A contractible loop therefore tests the curvature enclosed by the disk rather than a nontrivial element of \(\pi_1(T^2)\).  In the extracted local frame, the relevant infinitesimal quantity for the present purpose is the transverse component \((1-\Pi_{\rm loc})F_A\).  If its integral over the disk is nonzero after path ordering and cancellations are accounted for, the resulting holonomy can leave \(\U(2)\otimes\U(2)\) even though the loop is null-homotopic.

This also explains the orientation check.  For a fixed loop, reversing the orientation replaces the holonomy by its inverse, so \(U_-\simeq U_+^\dagger\) and subgroup-distance diagnostics are unchanged.  This statement is distinct from the BHZ comparison between \(C_+\sim(1,1)\) and \(C_-\sim(1,-1)\), which are different primitive cycles rather than opposite traversals of one cycle.  The strong path dependence is therefore the loop-level expression of the reduction problem: different loops test different transition functions of the extracted product structure.  A loop whose holonomy lies in the extracted local subgroup realizes a local gluing, whereas a loop whose holonomy leaves that subgroup realizes the entangling gluing obstruction described abstractly in Ref.~\cite{Ikeda2026}.

\begin{table}[H]
\centering
\begingroup
\caption{\textbf{Orientation check for a representative contractible circle.}  The same center-based loop \(\mathcal C_0(\rho)\), with \(\rho=0.53\), is traversed in the positive and reverse orientations.  The inverse consistency column reports \(\|U_- - U_+^\dagger\|_F\).}
\label{tab:contractible}
\scriptsize
\setlength{\tabcolsep}{4pt}
\begin{tabular}{llcccc}
\toprule
model & orientation & \(\Dloc\) & \(e_p\) & \(\max|\phi_j|\) & \(\|U_- - U_+^\dagger\|_F\) \\
\midrule
BHZ & \(+\) & \(6.77\times10^{-4}\) & \(1.02\times10^{-7}\) & \(3.60\times10^{-4}\) & \(5.71\times10^{-14}\) \\
BHZ & \(-\) & \(6.77\times10^{-4}\) & \(1.02\times10^{-7}\) & \(3.60\times10^{-4}\) & \(5.71\times10^{-14}\) \\
SSH & \(+\) & \(7.99\times10^{-4}\) & \(1.42\times10^{-7}\) & \(4.02\times10^{-4}\) & \(2.45\times10^{-14}\) \\
SSH & \(-\) & \(7.99\times10^{-4}\) & \(1.42\times10^{-7}\) & \(4.02\times10^{-4}\) & \(2.45\times10^{-14}\) \\
BBH & \(+\) & 0.1780 & \(6.97\times10^{-3}\) & 0.834 & \(2.74\times10^{-14}\) \\
BBH & \(-\) & 0.1780 & \(6.97\times10^{-3}\) & 0.834 & \(2.74\times10^{-14}\) \\
\bottomrule
\end{tabular}
\endgroup
\end{table}

\subsection*{Relation to Ref.~\cite{Ikeda2026}}

This relation is dimension-independent. Ref.~\cite{Ikeda2026} treats a general factorization type $\mathbf d=(d_1,\ldots,d_r)$. The present $G_{(2,2)}$ case is the smallest member of that family that can exhibit a nonlocal product-subgroup obstruction. Higher-dimensional boundary multiplets would replace the stabilizer $G_{(2,2)}$ by $G_{\mathbf d}$ without changing the reduction logic.

Ref.~\cite{Ikeda2026} asks a global existence question for subsystem structures on twisted projective families. Its mathematical core is a reduction theorem. For a chosen factorization type $d=(d_1,\dots,d_r)$, a global product-state locus exists precisely when the underlying $\mathrm{PGL}_n$-torsor reduces to the corresponding stabilizer $G_d$. The resulting subsystem structures are parametrized by $P/G_d$ and can be realized as a relative Hilbert-scheme locus. Here we fix a spectrally isolated rank-four boundary quartet and one observable pair $(O_A,O_B)$ for each model, and ask a loopwise question instead. Which microscopic holonomies stay inside the extracted local subgroup, and which leave it? In that sense, the condensed-matter result is a loopwise counterpart of Ref.~\cite{Ikeda2026}, not a reformulation of its global existence theorem.

This obstruction should be distinguished from a Brauer-class obstruction. The present quartet is realized inside a globally defined microscopic Hilbert space. After projection to the isolated four-state bundle, the associated algebra of endomorphisms is of the form $\operatorname{End}(E)$ and is therefore Brauer-trivial. The relevant obstruction is the finer subsystem-reduction obstruction identified in Ref.~\cite{Ikeda2026}. Even when the Brauer class is zero, a projective torsor or its gluing functions may fail to reduce to the stabilizer $G_{(2,2)}$. In the present unitary setting, this becomes the loopwise subgroup-membership test $U(\mathcal C)\in \U(2)\otimes\U(2)$. If the test fails, the loop glues the extracted product structure by an entangling element, but it does not define a new Brauer class.

For the present rank-four setting with factorization type $(2,2)$, the stabilizer in Ref.~\cite{Ikeda2026} is the projective counterpart of local two-qubit transformations. Transition functions of an abstract projective torsor are replaced here by holonomies of an isolated boundary quartet. Noncontractible cycles probe global gluing data of the quartet connection. Contractible circles probe the same reduction question infinitesimally through the nonlocal component of the non-Abelian Berry curvature. In both cases the point is not whether the loop is topologically nontrivial. The point is whether the associated gluing operator preserves the extracted product structure.

\FloatBarrier
\subsection*{Observable robustness and controls}

\begin{figure}[H]
  \centering
  \includegraphics[width=\textwidth]{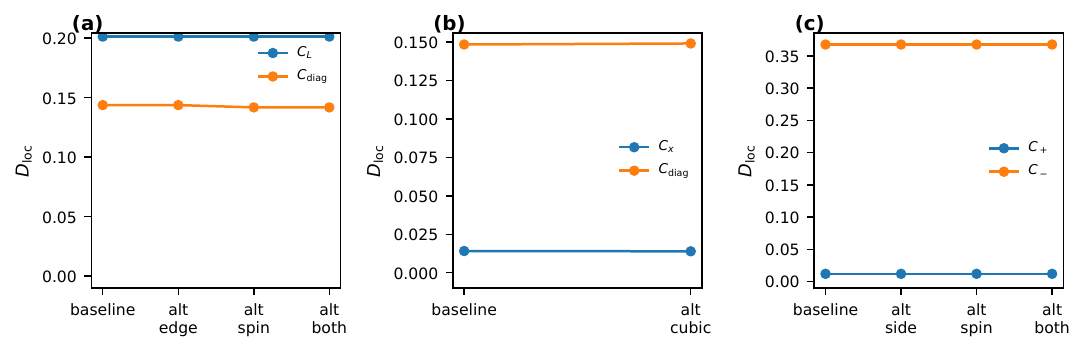}
  \caption{Observable-choice robustness. (a) SSH under coarse edge and rotated spin-like observables. (b) BBH under cubic coarse position observables. (c) BHZ under side-smoothing and rotated spin-like observables. In every case the qualitative loop-dependent pattern remains unchanged.}
  \label{fig:robust}
\end{figure}

Because the construction starts from two compressed observables, robustness under observable choice must be checked. We therefore varied the coarse edge and spin-like observables in all three models. The results are summarized in \Cref{fig:robust}. In SSH the left loop remains at $0.20$ while the diagonal loop stays near $0.14$ under rotated spin-like observables. In BBH the axis loop stays at $0.01$ and the diagonal loop near $0.15$ under cubic coarse position observables. In BHZ the co-rotating and counter-rotating distances are unchanged within plotting resolution under both side-smoothing and rotated spin-like observables. The loop-dependent pattern is therefore not an artifact of any particular observable choice. Because the extraction uses a fixed order, we also exchanged the roles of $O_A$ and $O_B$ on the representative loops. \Cref{tab:orderswap} shows that this changes the numbers only slightly: SSH and BHZ keep the hierarchy among almost local, intermediate, and entangling loops, and BBH keeps the axis-versus-mixed separation.

\begin{table}[H]
\centering
\caption{\textbf{Representative order-swap robustness.} Distances are recomputed after exchanging $O_A$ and $O_B$ in the sequential extraction. The qualitative hierarchy is unchanged.}
\label{tab:orderswap}
\small
\setlength{\tabcolsep}{5pt}
\begin{tabular}{llcc}
\toprule
model & loop family & original order & swapped order \\
\midrule
SSH & $C_L$, $C_{\rm diag}$, $C_{\rm anti}$ & 0.20, 0.14, 0.38 & 0.20, 0.14, 0.38 \\
BHZ & $C_+$, $C_T$, $C_-$ & 0.01, 0.18, 0.37 & 0.01, 0.18, 0.37 \\
BBH & $C_x$, $C_y$, $C_{\rm diag}$ & 0.01, 0.008, 0.15 & 0.01, 0.008, 0.15 \\
\bottomrule
\end{tabular}
\normalsize
\end{table}

\begin{figure}[t]
  \centering
  \includegraphics[width=\textwidth]{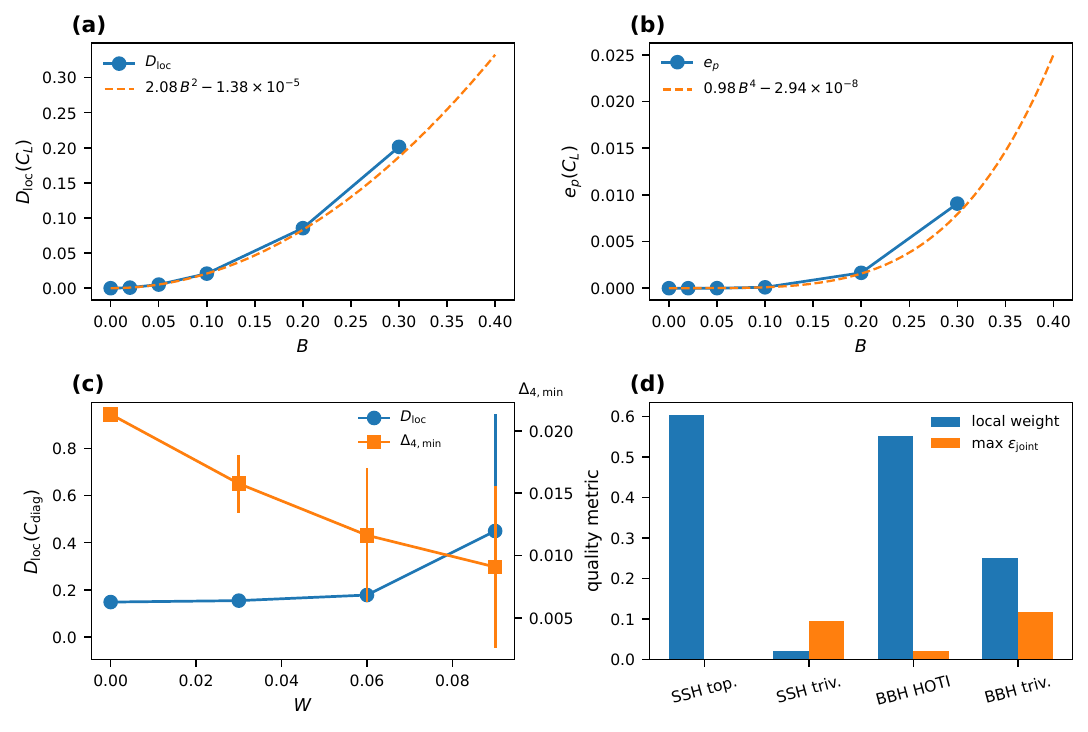}
  \caption{Additional controls. (a,b) SSH small-$B$ scan. The response follows quadratic scaling in $\Dloc$ and quartic scaling in $\ep$ for weak driving. (c) BBH disorder scaling: a controlled regime persists for weak disorder before quartet degradation sets in. (d) Trivial dimerization and non-HOTI controls fail the quartet-quality prerequisites, showing that the benchmark requires a localized topological multiplet.}
  \label{fig:controls}
\end{figure}

The small-$B$ SSH control shows a perturbative onset with the microscopic driving. For weak $B$,
\begin{equation}
\Dloc(C_L)\propto B^2,
\qquad
\ep(C_L)\propto B^4.
\end{equation}
This indicates a perturbative origin for the local-subgroup obstruction rather than a threshold artifact.

The corrected BBH disorder protocol exhibits a controlled weak-disorder regime rather than broad robustness. In that regime the average diagonal-loop distance remains near $0.16$, while at larger disorder the variance grows and the quartet gap is reduced. The appropriate interpretation is therefore stability only before quartet degradation sets in.

Finally, trivial-phase controls show that a topological low-energy multiplet is a necessary prerequisite for the present benchmark. On the trivial SSH side the edge weight collapses to approximately $0.02$, $\max\epsjoint$ rises to approximately $0.09$, and the minimal gap drops to approximately $0.03$. On the non-HOTI BBH side the corner weight falls to approximately $0.25$ and $\max\epsjoint$ rises to approximately $0.12$. A matching BHZ trivial control at $M=5$ makes the same point in BHZ itself: at fixed $k_x=0$, the minimal quartet gap drops to $0.16$, the outer-two-row weight to $0.19$, and $\max\epsjoint$ rises to $0.91$. In all three cases the four-state truncation ceases to be a reliable edge or corner benchmark before any loop-level interpretation becomes reliable.

\clearpage
\bibliographystyle{JHEP}
\bibliography{refs_v4}

\end{document}